# THE SIMPLE NONPOLAR CONTINUUM MEDIA.
# PART II. THE CONSTITUTIVE EQUATIONS.
# LINEAR STRUCTURES.

V.O. Bytev

*10 Semakova Street, Tyumen State University,Tyumen, Russia, 625003, e-mail: vbytev@utmn.ru*
*Head of Appl. Math. Dep. Tyumen State University*

To C. Truesdell
"… the supposed is rarely separated from what is to be proved and the last instance in every particular case of doubt is the model of a matter consisting of small solid balls"
C. Truesdell (see, e.g. [1])

Abstract. A method for obtaining the constitutive equations for describing the simple nonpolar continuum media is given. The method is based on the theory of Lie groups and on group properties of the differential equations.

As it is well known, the idea of using the group of continuum transformations in mathematical modeling is rather old, but its technical implementation was not clear. The success of the group approach in crystal physics was conditioned by the fact that all the discrete groups characterizing any crystal were known, whereas the usage of groups of continuum transformations faced an insurmountable obstacle, i.e., the absence of the results of Shubnikov's classification types.

These difficulties can be eliminated in the case where the problem of group classification of differential consequences of conservation laws is solved. The problem of group classification is the generalization of the problem of the basic group calculation and is of interest from the view-point of applications, since its solution allows us to select the most interesting values and forms of the experimentally determined quantities and values. The problem's statement runs as follows [2]. Let the system $S$ of differential equations contain some parameters or functions, the arbitrary ones specialization of these functions results in specialization of the initial system $S_0$. In case $S$ assumed some group $g$ independently from a specialization of an arbitrary element, this group $g$ is assumed by any specialization. It is obvious that specializations of arbitrary functions are possible when group $g_0 \supset g$. This very principle, i.e., the partial ordering in respect to inclusion, is taken as a basis in the classification of all specializations of the arbitrary functions constituting the initial system $S$. In brief, the search of all specializations $S_0$ (where the basic group $g \subset g_0$) for the system $S$ of differential equations, including the arbitrary element and assuming the independent basic group $g$.

Let us set the following equations as the system $S$:

$$\rho_t + div(\rho \vec{u}) = 0 \tag{1.1}$$

$$\rho\left[\vec{u}_t + \vec{u}\nabla\vec{u}\right] - div\,\Pi + \nabla p = 0 \tag{1.2}$$

$$p_t + \vec{u}\nabla p + G\,div\,\vec{u} + H\Phi = 0 \tag{1.3}$$

Here, $\rho$ is density of the continuum medium considered; $p$ is the hydrostatic (equilibrium) pressure; $\vec{u}$ is the velocity vector; $\Pi$ is the tensor of viscous tension;

$$G(p,\rho) = -\rho\,S_\rho/S_p\,; \qquad H(p,\rho) = -\frac{1}{\rho\,\Theta\,S_p}\,; \qquad S = S(p,\rho)$$

is the entropy; $\Theta = \Theta(p,\rho)$ is absolute temperature; and $\Phi = (\Pi : \nabla \vec{u})$ is the dissipative function. Allow $\Pi$ to be a symmetric tensor, depending only on $\nabla \vec{u}$.

The traditional approach is the overlapping of the requirement for tensor $\Pi$ invariance relative to the action of the $SO_3$ group. Then the classical notion of it in this case follows

$$\Pi = \varphi_0 I + \varphi_1 D + \varphi_2 D^2.$$

Here, we get $I$ the single tensor; $D$ the tensor of the deformation velocities – the symmetric part of the tensor $\nabla \vec{u}$; and $\varphi_0, \varphi_1, \varphi_2$ the functions of invariants $D$. Such an approach is justified by the supposition of isotropy of a considered continuum medium. But still, it is a weak motivation for making such a strong assertion. Really, as it had been shown the group equivalence for system $S$ do not contain $SO_3$ [3].

Let us demand the invariance of a variety (set up by the system $S$) in respect to the action of the group of continuum transformations. It results in the following [4]. Let the general operator of infinitesimal transformation be

$$X = \xi^0 \frac{\partial}{\partial t} + \eta^0 \frac{\partial}{\partial p} + \xi^i \frac{\partial}{\partial x^i} + \eta^k \frac{\partial}{\partial u^k} + \eta^4 \frac{\partial}{\partial \rho},$$

where

$$\xi^n = \xi^n(t, \vec{x}, \vec{u}, p, \rho); \quad \eta^m = \eta^m(t, \vec{x}, \vec{u}, p, \rho); \quad n = 0 \div N; \quad m = 0 \div N+1.$$

Summarizing from *1* to *N* is performed by the repeated indices. Then

$$\begin{aligned}
\xi^0 &= a_0 t^2 + a_2 t + d; \\
\xi^i &= a_0 t x^i + a_{ik} x^k + a_5 x^i + a_{i0} t + b_i; \\
a_{ik} &= \begin{cases} -a_{ki}, i \neq k \\ 0, i = k \end{cases} \quad (i, k = 1 \div N); \\
\eta^i &= -(a_0 t - a_5) u^i + a_{ik} u^k + a_0 x^i + a_{i0}; \\
\eta^4 &= (e - 2a_5 - N a_0 t) \rho; \\
\eta^0 &= [e - (N+2) a_0 t] p + r,
\end{aligned} \qquad (1.4)$$

Were $a_0, a_2, d, a_{ik}, a_{i0}, a_5, b_i, e, r$; are the group constants with the corresponding infinitesimal operators

$$a_0 : Z_0 = t^2 \frac{\partial}{\partial t} + t x^i \frac{\partial}{\partial x^i} - (t u^i - x^i) \frac{\partial}{\partial u^i} - (N+2) t p \frac{\partial}{\partial p} - N t \rho \frac{\partial}{\partial \rho};$$

$$a_2 : Z_1 = t \frac{\partial}{\partial t} + x^i \frac{\partial}{\partial x^i};$$

$$a_5 : Z_2 = x^i \frac{\partial}{\partial x^i} + u^i \frac{\partial}{\partial u^i} - 2\rho \frac{\partial}{\partial \rho};$$

$$e : Z_3 = p \frac{\partial}{\partial p} + \rho \frac{\partial}{\partial \rho};$$

$$d : X_0 = \frac{\partial}{\partial t}, \quad b_i : X_i = \frac{\partial}{\partial x^i}, \quad (i = 1 \div N);$$

$$a_{i0} : Y_i = t\frac{\partial}{\partial x^i} + \frac{\partial}{\partial u^i};$$

$$a_{ik} : X_{lk} = x^k \frac{\partial}{\partial x^l} - x^l \frac{\partial}{\partial x^k} + u^k \frac{\partial}{\partial u^l} - u^l \frac{\partial}{\partial u^k}; \quad l < k; \; l,k = 1 \div N;$$

$$r : S = \frac{\partial}{\partial p}; \quad N = 1, 2, 3.$$

(1.5)

Besides, the equations playing the role of restrictions overlapping the width of the group assumed by the system $S$, have been obtained as a result of the solution of the problem of group analysis. These equations are written as

$$a_0 \left[ (N+2)(pG_p - G) + N\rho G_\rho \right] = 0;$$
$$a_0 \left[ (N+2)p - NG \right] = 0;$$
$$a_0 \left[ (N+2)pH_p + N\rho H_\rho \right] = 0;$$
$$a_0 \frac{(N+2)}{2} \Pi^{kt} = a_0 p_j^i \frac{\partial \Pi^{kt}}{\partial p_j^i};$$
$$a_0 H \, tr\Pi = 0;$$
$$a_0 \sum_{i=1}^N \frac{\partial \Pi^{kt}}{\partial p_i^i} = 0;$$
$$e(\rho G_\rho + pG_p - G) + rG_p - 2a_5 \rho G_\rho = 0;$$
$$e(\rho H_\rho + pH_p) + rH_p - 2a_5 \rho H_\rho = 0;$$
$$a_{ki}\Pi^{it} + a_{ti}\Pi^{ki} + e\Pi^{kt} + a_2 p_s^l \frac{\partial \Pi^{kt}}{\partial p_s^l} - a_{lj} \frac{\partial \Pi^{kt}}{\partial p_s^l} p_s^j + a_{js} \frac{\partial \Pi^{kt}}{\partial p_s^l} p_j^l = 0.$$

(1.6)

The notion $p_j^i \equiv \frac{\partial u^i}{\partial x^j}$ is introduced.

Proceeding from this system of equations, it is not difficult to conclude that the assumed equations under the arbitrary $\Pi$, $G$, $H$ allow the following operators

$$X_0, Y_i, X_i, \quad (i = 1 \div N),$$

(1.7)

it means that the variety $S$ under the arbitrary $\Pi$, $G$ and $H$ is invariant only with regards to the action of the group of transitions and of the Galilei transitions. Hereafter, we will denote this group as $\Gamma_0$. Further, the maximal group of continuum transformations assumed in the sense of Lee's equations setting up the variety $S$, corresponds to the case

$$\Pi \equiv 0, \quad G = \frac{N+2}{N} p.$$

(1.8)

But in this situation we have the other manifolds and the other system of differential equations. Namely, we have the system, which describe any moving of gas and the group properties of them are well known [2, 5].

Now the problem is to determine all the possible types of relations between group constants included in the described system of equations and, consequently, to get the systems of differential equations for the components of viscous stress tensor $\Pi$ and for the functions $G$ and $H$. Thus, we not only get the general representations on the stress tensor and the corresponding state equations, but the more wide groups of continuum transformations which are assumed by the motion equations in every particular case of tensor $\Pi$ and functions $G$, $H$ specialization as well. In other words, the group $\Gamma_i \supset \Gamma_0$ corresponds to every three elements $\Pi_i$, $G_i$, $H_i$.

*Remark.* No other models, except the ones obtained, exist within the framework of this approach (i.e., group classification) and the suppositions made.

The pairs of functions $\{G, H\}$ which deserve to be [4]

$$\left\{pG(p^{1-2m}\rho^{-1}); H(p^{1-2m}\rho^{-1})\right\}; \quad \left\{G(\rho e^{2p}); H(\rho e^{2p})\right\};$$
$$\left\{G(\rho); H(\rho)\right\}; \quad \{G_0 p^m; H_0\}; \quad \{G(p); H(p)\}; \quad (1.9)$$
$$\{1; H_0\}; \quad \{0; H(\rho)\}; \quad \{0; H_0\}; \quad \{\gamma p; H_0\}$$

$G_0$, $H_0$ are constants. If $a_0 \neq 0$, then only following pairs are to be reviewed

$$\left\{\frac{N+2}{N}p; H_0\right\}; \quad \left\{\frac{N+2}{N}p; H(p^{N/(N+2)}\rho^{-1})\right\}. \quad (1.10)$$

**Linear Structures**

Let

$$\Pi^{11} = a_s^l p_l^s; \quad \Pi^{12} = \Pi^{21} = b_s^l p_l^s; \quad \Pi^{13} = \Pi^{31} = c_s^l p_l^s;$$
$$\Pi^{22} = d_s^l p_l^s; \quad \Pi^{23} = \Pi^{32} = e_s^l p_l^s; \quad \Pi^{33} = f_s^l p_l^s. \quad (2.1)$$

Substituting (2.1) into (1.6), we obtain $a_0 = 0$, and the system of equations (1.6) defining the range of the assumed group will be simplified and become

$$e(\rho G_\rho + p G_p - G) + r G_p - 2a_5 \rho G_\rho = 0; \quad (2.2)$$
$$e(\rho H_\rho + p H_p) + r H_p - 2a_5 \rho H_\rho = 0; \quad (2.3)$$
$$(e + a_2)\Pi^{11} + a_{12}(L_1\Pi^{11} + 2\Pi^{12}) + a_{13}(L_2\Pi^{11} + 2\Pi^{13}) + a_{12}L_3\Pi^{11} = 0; \quad (2.4)$$
$$(e + a_2)\Pi^{12} + a_{12}(L_1\Pi^{12} + \Pi^{22} - \Pi^{11}) +$$
$$+ a_{13}(L_2\Pi^{12} + \Pi^{23}) + a_{23}(L_3\Pi^{12} + \Pi^{13}) = 0; \quad (2.5)$$
$$(e + a_2)\Pi^{13} + a_{12}(L_1\Pi^{13} + \Pi^{23}) +$$
$$+ a_{13}(L_2\Pi^{13} + \Pi^{33} - \Pi^{11}) + a_{23}(L_3\Pi^{13} - \Pi^{12}) = 0; \quad (2.6)$$
$$(e + a_2)\Pi^{22} + a_{12}(L_1\Pi^{22} - 2\Pi^{12}) + a_{13}L_2\Pi^{22} + a_{23}(L_3\Pi^{22} + 2\Pi^{23}) = 0; \quad (2.7)$$

$$(e + a_2)\Pi^{23} + a_{12}(L_1\Pi^{23} - \Pi^{13}) + \qquad (2.8)$$
$$+ a_{13}(L_2\Pi^{23} - \Pi^{12}) + a_{23}(L_3\Pi^{23} + \Pi^{33} - \Pi^{22}) = 0;$$
$$(e + a_2)\Pi^{33} + a_{12}L_1\Pi^{33} + a_{13}(L_2\Pi^{33} - 2\Pi^{13}) + a_{23}(L_3\Pi^{33} - 2\Pi^{23}) = 0. \qquad (2.9)$$

Here $L_1$, $L_2$, $L_3$, $L_4$ are the linear differential operators of view

$$\begin{aligned}
L_1 &= (p_1^1 - p_2^2)\left(\frac{\partial}{\partial p_2^1} + \frac{\partial}{\partial p_1^2}\right) + (p_2^1 - p_1^2)\left(\frac{\partial}{\partial p_2^2} + \frac{\partial}{\partial p_1^1}\right) + \\
&\quad + p_3^1\frac{\partial}{\partial p_3^2} - p_3^2\frac{\partial}{\partial p_3^1} + p_1^3\frac{\partial}{\partial p_2^3} - p_2^3\frac{\partial}{\partial p_1^3}; \\
L_2 &= (p_1^1 - p_3^3)\left(\frac{\partial}{\partial p_1^3} + \frac{\partial}{\partial p_3^2}\right) + (p_3^1 - p_1^3)\left(\frac{\partial}{\partial p_3^3} + \frac{\partial}{\partial p_1^1}\right) + \\
&\quad + p_2^1\frac{\partial}{\partial p_2^3} - p_2^3\frac{\partial}{\partial p_2^1} + p_1^2\frac{\partial}{\partial p_3^2} - p_3^2\frac{\partial}{\partial p_1^2}; \\
L_3 &= (p_2^2 - p_3^3)\left(\frac{\partial}{\partial p_2^3} + \frac{\partial}{\partial p_3^2}\right) + (p_3^2 - p_2^3)\left(\frac{\partial}{\partial p_2^2} + \frac{\partial}{\partial p_1^1}\right) + \\
&\quad + p_2^1\frac{\partial}{\partial p_3^1} - p_3^1\frac{\partial}{\partial p_2^1} + p_1^2\frac{\partial}{\partial p_1^3} - p_1^3\frac{\partial}{\partial p_1^2}; \\
L_4 &= \sum_{l=1}^{N}\sum_{s=1}^{N} p_s^l \frac{\partial}{\partial p_l^s}.
\end{aligned} \qquad (2.10)$$

For the classification of the state equations which has already been accomplished, it becomes necessary to review only the following cases of possible connections between constants $e$, $a_{12}$, $a_{13}$, $a_{23}$, $a_2$.

$$a_{12} = a_{13} = a_{23} = 0,$$

then $e = -a_2$. The condition of the nonnegativeness of the dissipative function $\Phi$ provides restrictions of a nonequality type for the coefficients in the ratio (2.1) which are arbitrary in other respects. Let $a_{12}$, $a_{13}$, $a_{23}$ not to be equal to zero. In this case, we have a simple linear model. Note that in this particular case, $e + a_2 = 0$ as well. It is a consequence of the fact that $L_1$, $L_2$, $L_3$, $L_4$ from the Lee algebra.

Let us consider the case with rotary symmetry motions. For instance, let $a_{13} = a_{23} = 0$, $a_{12} \neq 0$, $e + a_2 = 0$. Thus, the system of equations (2.2)-(2.9) with regard to be obtained connections between group constants will be given by

$$e(\rho G_\rho + p G_p - G) + r G_p - 2a_5 \rho G_\rho = 0; \qquad (2.11)$$
$$e(\rho H_\rho + p H_p) + r H_p - 2a_5 \rho H_\rho = 0; \qquad (2.12)$$
$$L_1\Pi^{11} + 2\Pi^{12} = 0; \qquad (2.13)$$
$$L_1\Pi^{12} + \Pi^{22} - \Pi^{11} = 0; \qquad (2.14)$$
$$L_1\Pi^{13} + \Pi^{23} = 0; \qquad (2.15)$$
$$L_1\Pi^{22} - 2\Pi^{12} = 0; \qquad (2.16)$$
$$L_1\Pi^{23} - \Pi^{13} = 0; \qquad (2.17)$$

$$L_1\Pi^{33} = 0. \tag{2.18}$$

The pair of equations (2.11) and (2.12) has already been investigated [4]. The general solution of the system (2.13)-(2.18) will be represented as follows (with the nonnegativeness of the dissipative function taken into account)

$$\begin{aligned}
\Pi^{11} &= 2au_x + b(u_x + v_y) + c(u_y + v_x) + 2ew_z; \\
\Pi^{12} &= -c(u_x - v_y) + a(u_y + v_x); \\
\Pi^{22} &= b(u_x + v_y) + 2av_y - c(u_y + v_x) + 2ew_z; \\
\Pi^{33} &= \varepsilon w_z + 2f(u_x + v_y); \\
\Pi^{13} &= \gamma_1(w_x + u_z) + \gamma_2(w_y + v_z); \\
\Pi^{23} &= -\gamma_2(w_x + u_z) + \gamma_1(w_y + v_z).
\end{aligned} \tag{2.19}$$

Let us calculate the dissipative function $\Phi = (\Pi : \nabla \vec{u})$.
We obtain

$$\Phi = \left(\sqrt{a+b}\,(u_x + v_y) + \frac{e+f}{\sqrt{a+b}}\,w_z\right)^2 + a(u_x - v_y)^2 + \\
+ a(u_y + v_x)^2 + \gamma_1(w_x + u_z)^2 + \gamma_1(w_y + v_z)^2 + \left(\varepsilon - \frac{(e+f)^2}{a+b}\right)w_z^2 > 0, \tag{2.20}$$

with

$$a \geq 0, \quad a+b > 0, \quad \gamma_1 \geq 0, \quad \varepsilon \geq (e+f)^2/(a+b); \tag{2.21}$$
$c, \gamma_2, e, f$ are arbitrary.

The main result is the theorem 1.

Theorem 1.  Let the system (1.1)-(1.3) admit of only one operator of rotation. Then the most general representation of viscous stress tensor $\Pi$ have the components (2.19).

Corollary 1.  If $a = b = \gamma_1 = 0$, then $\Phi = \varepsilon w_z^2 > 0$, if and only if $e + f = 0$.

Corollary 2.  If $a = b = \gamma_1 = \varepsilon = 0$, and $e + f = 0$, than there are processes with viscosities $c, \gamma_2, e, f$ but without dissipation.

Corollary 3.  If $a + b = 0$ then $e + f = 0$, and
$$\Phi = 4a(u_x - v_y)^2 + a(u_y + v_x)^2 + \gamma_1(w_x + u_z)^2 + \gamma_1(w_y + v_z)^2 + \varepsilon w_z^2;$$
$a \geq 0, \quad \gamma_1 \geq 0, \quad \varepsilon \geq 0.$

Corollary 4.  Let $N = 2$ then
$$\begin{aligned}
\Pi^{11} &= 2au_x + b(u_x + v_y) + c(u_y + v_x); \\
\Pi^{12} &= -c(u_x - v_y) + a(u_y + v_x); \\
\Pi^{22} &= b(u_x + v_y) + 2av_y - c(u_y + v_x);
\end{aligned}$$

or more short
$$\Pi = MD_0 + I(\mu + b)\,div\,\vec{u}.$$

Here $a = \mu;\quad c = \mu_0;\quad M = \begin{pmatrix} \mu & \mu_0 \\ -\mu_0 & \mu \end{pmatrix};\quad D_0 = \begin{pmatrix} u_x - v_y & u_y + v_x \\ u_y + v_x & v_y - u_x \end{pmatrix};$

and
$$\Phi = 4(\mu + b)(div\,\vec{u})^2 + 4\mu(u_y - v_x)^2 + \mu(u_y + v_x)^2 \geq 0;$$
$\mu \geq 0,\quad \mu + b \geq 0,\quad \mu_0$ is arbitrary.

Conclusion. In such a way we can assume that the reason of nonstability and turbulence of flow into a pipes is possible for an existence and an interaction between dissipative and nondissipative processes.

Example. Singling out the deviator of tensor $\Pi$ in the presented notation, we find that it almost fully coincides with the deviator of the stress tensor found in [5] by means of the methods of statistical physics for the description of quasi-neutral plasma motion in the one-liquid approximation. Next, we define the structure of this tensor

$$\begin{aligned}
\Pi^{11} &= \eta\left\{b_2' e_{11} + \frac{1}{2}(b_2' - 1)e_{33} - 2\omega_2\tau_2 b_2'' e_{12}\right\}; \\
\Pi^{22} &= \eta\left\{b_2' e_{22} + \frac{1}{2}(b_2' - 1)e_{33} + 2\omega_2\tau_2 b_2'' e_{12}\right\}; \\
\Pi^{33} &= \eta e_{33}; \\
\Pi^{12} &= \Pi^{21} = \eta\left\{b_2' e_{12} - \omega_2\tau_2 b_2''(e_{11} - e_{22})\right\}; \\
\Pi^{13} &= \Pi^{31} = \eta\left\{b_1' e_{13} - \omega_2\tau_1 b_1'' e_{23}\right\}; \\
\Pi^{23} &= \Pi^{32} = \eta\left\{b_1' e_{23} + \omega_2\tau_2 b_1'' e_{13}\right\};
\end{aligned} \quad (2.22)$$

Here
$$e_{ik} = p_k^i + p_i^k - \frac{2}{3}\delta_{ik}(p_1^1 + p_2^2 + p_3^3);$$
and $\eta$, $b_1'$, $b_1''$, $b_2'$, $b_2''$ are kinetic coefficients.

To make comparison (2.19) and (2.22) we obtain:

$$a = \eta b_2';\quad a + b = \frac{1}{3}\eta;\quad e = -\frac{1}{3}\eta;$$
$$c = -2\eta\omega_2\tau_2 b_2'';\quad f = -\frac{1}{3}\eta;\quad \varepsilon = \frac{4}{3}\eta;$$
$$\gamma_1 = \eta b_1';\quad \gamma_2 = -\eta\omega_2\tau_1 b_1'' = -\eta\omega_2\tau_2 b_1''.$$

From this equality follows

$$\tau_1 = \tau_2.$$

The next part will be devoted to the theory of elasticity.

*References*

[1]  Truesdell, C.: 1966, "Six Lectures on Modern Natural Philosophy", Springer Science-Verlag, New York.

[2]  Ovsyannikov, L. V.: 1978, "The Group Analysis of Differential Equations", Nauka, Moscow (in Russian). English translation, Ames, W.F., Ed., published by Academic Press, New York, 1982.

[3]  Bytev, V.O.: 2009, "The Simple Nonpolar Continuum Media". Part I. The equivalence transformation. (archived article, www.arxiv.org)

[4]  Bytev, V.O.: 1989, "Building of Mathematical Models of Continuum Media on the Basis of the Invariance Principle". Acta Appl. Math. 16: p.117-142, Kluwer Acad. Publ. Netherlands.

[5]  Gubanov, E. and Lunkin, I.: 1960, "The Magnetoplasmadynamics Equations" Jorn. of Techn. Phys. Vol.XXX №9 p. 1046-1053.